\documentclass[prl,twocolumn,groupedaddress]{revtex4}
\usepackage{graphics,amssymb}

\pdfoutput=1 

\newcommand{\fig}[5]
{
\begin{figure}[#4]
\begin{center}
\resizebox{#5 \columnwidth}{!}{\includegraphics{#1}}
\caption{\label{#3}#2}
\end{center}    
\end{figure}
}

\newcommand{\figstar}[4]
{
\begin{figure*}[#4]
\begin{center}
\resizebox{1.90 \columnwidth}{!}{\includegraphics{#1}}
\caption{\label{#3}#2}          
\end{center}        
\end{figure*}
}

\begin{document}

\title{Relationship between vibrations and dynamical heterogeneity in a model glass former: extended soft modes but local relaxation}
\author{Douglas J. Ashton and Juan P. Garrahan}

\affiliation{School of Physics and Astronomy, University of Nottingham, Nottingham, NG7 2RD, U.K.}

\begin{abstract}
We study the relation between short-time vibrational modes and long-time relaxational dynamics in a kinetically constrained lattice gas with harmonic interactions between neighbouring particles.   We find a correlation between the location of the low (high) frequency vibrational modes and regions of high (low) propensity for motion.  This is similar to what was observed in continuous force systems, but our interpretation is different: in our case relaxation is due to localised excitations which propagate through the system; these localised excitations act as background disorder for the elastic network, giving rise to anomalous vibrational modes.  Our results show that a correlation between spatially extended low frequency modes and high propensity regions {\em does not} imply that relaxational dynamics originates in extended soft modes.   We consider other measures of elastic heterogeneity, such as non-affine displacement fields and mode localisation lengths, and discuss implications of our results to interpretations of dynamic heterogeneity more generally.
\end{abstract}

\maketitle

Two of the central features of glass forming systems are, on the one hand, the increasingly fluctuating, heterogenous dynamics \cite{reviewsDH} that accompanies the growth of relaxational timescales in supercooled liquids \cite{reviews}, and on the other, the anomalies in their vibrational spectrum when considered as amorphous solids \cite{reviewsBP,Jamming}.  While the former pertains to structural relaxation on long times (say of the order of seconds for a liquid near the experimental glass transition temperature), and the latter to vibrational motion at high frequency (say in the THz for the Boson peak  \cite{reviewsBP} of glasses), an important question is whether these two characteristic aspects of the dynamics of glass formers are related, and in particular whether there is a common structural origin for both of them.  

Interesting recent studies of systems of hard spheres \cite{Wyart1}, hard disks \cite{Wyart2}, and soft disks \cite{Harrowell} have revealed that the spatial localization of the anomalous low frequency modes of vibrations around either coarse-grained configurations in meta-basins \cite{Wyart1,Wyart2}, or around inherent-structures \cite{Harrowell}, give a good indication of the spatial distribution of structural relaxation at much longer times.  It has been argued \cite{Wyart1,Wyart2} that this correspondence is causal \cite{Harrowell}, in that it is these soft vibrational modes which provide the underlying structural mechanism to long time relaxation.

\fig{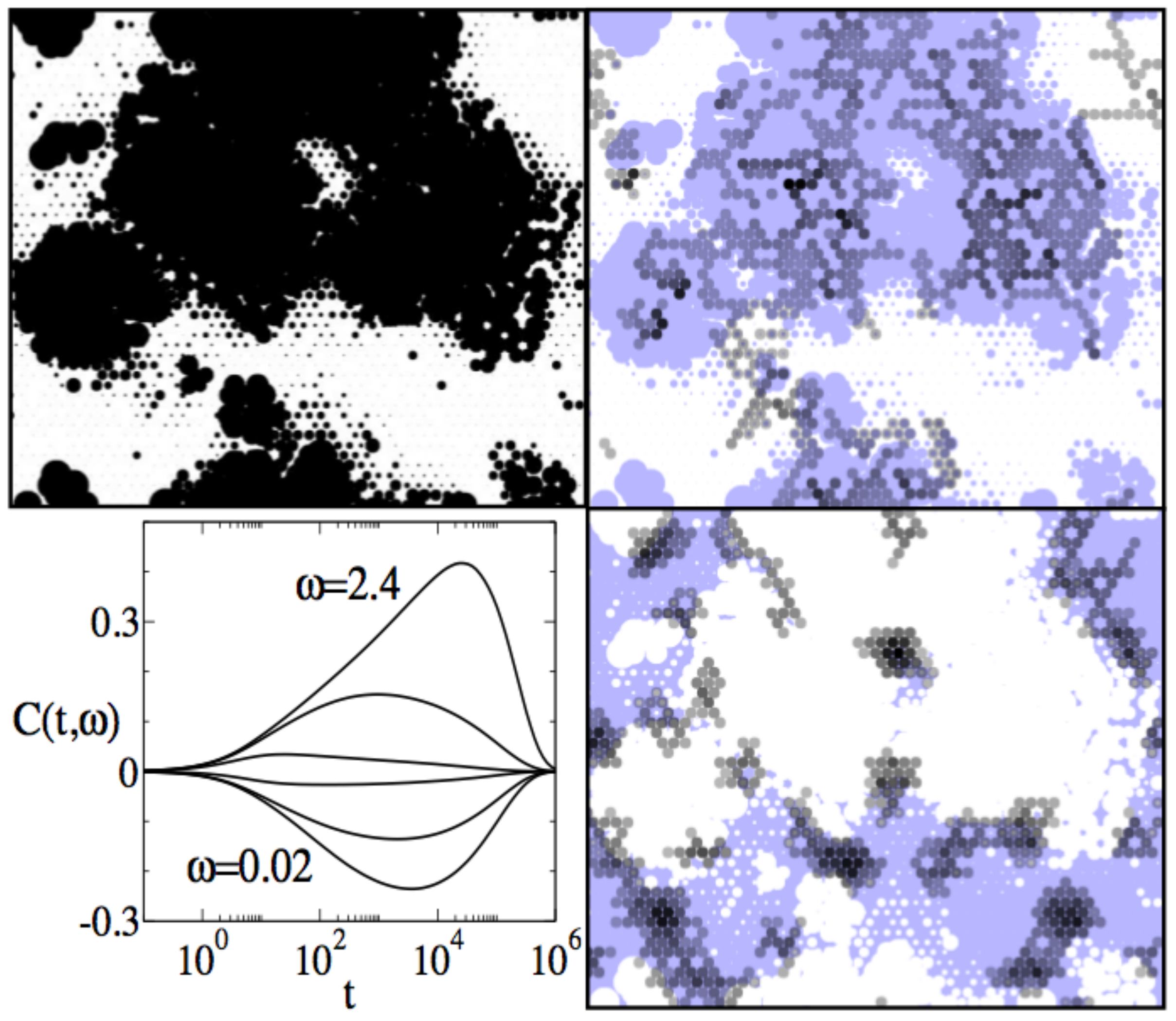}{Clockwise from top left: (a) Propensity map of the (2)-TLG at density $\rho=0.75$ averaged over $100$ trajectories.  Black circles indicate the isoconfigurational average distance travelled after time $t=5 \times 10^3$.  (b) The average participation, $\nu_i(\omega)$ of the lowest frequency modes, $\omega < 0.1$, and (c) the highest frequency modes, $\omega > 2.4$, of the central force network. (d) The connected correlation, $C(t,\omega)$, between isoconfigurational persistence, $p_{i}^{\mathrm{IC}}(t)$, and the participation, $\nu_i(\omega)$, as defined in the text.  From bottom to top the frequencies are $\omega=0.02,0.1,1,2,2.3,2.4$.}{fig:one}{hb}{.98}

In this paper we address this problem by studying spatial correlation between structural relaxation and vibrational modes in suitably generalized kinetically constrained models of glasses \cite{Ritort-Sollich}.  Our main results are illustrated in Fig.\ 1.  The top left panel is a``propensity'' map \cite{Harrowell2,Lester} of a two-vacancy assisted lattice gas, or (2)-TLG \cite{Jackle}, showing the average spatial distribution of structural relaxation starting from a typical configuration (details are given below).  The right hand panels show the location of the lowest frequency modes (top) and highest frequency modes (bottom) of the elastic network obtained from the same configuration when neighbouring particles interact via harmonic springs.  There is a clear spatial correlation between low frequency modes and high propensity for relaxation, and high frequency modes and low propensity for relaxation, as in atomistic models \cite{Wyart1,Wyart2,Harrowell}.  In our model, however, relaxation is via the propagation of localized clusters of vacancies \cite{Toninelli,Pan,Lester}, and does not emanate from the soft vibrational modes.  In fact, it is precisely the presence of these localized defects in the elastic network that gives rise to the anomalous soft modes.   We conclude that soft vibrational modes do indeed provide a good indicator of long time dynamic heterogeneity,  but this correlation does not imply they are the underlying mechanism for long time relaxational dynamics in glass formers in general.

\fig{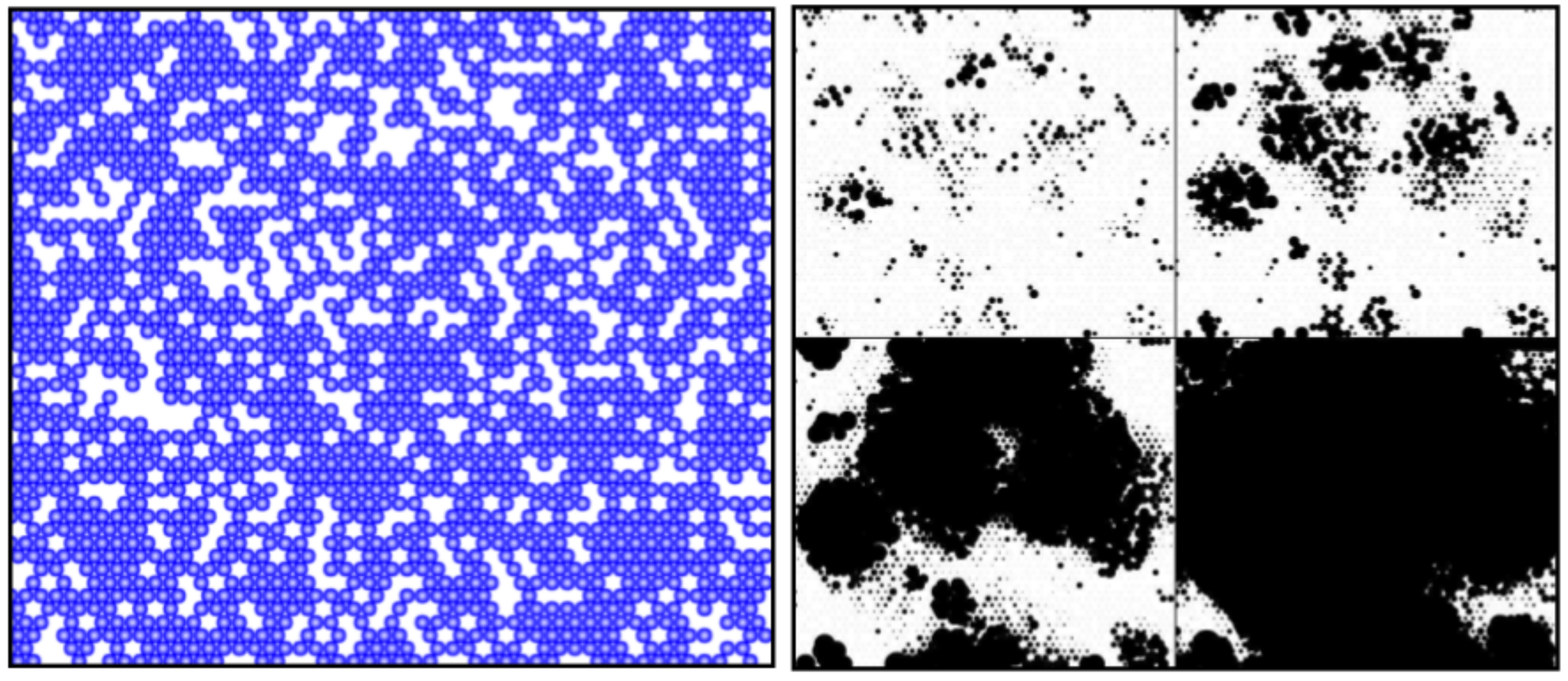}{Propensity maps for a (2)-TLG at density $\rho=0.75$ (second to fifth panels) for the same configuration as Fig.\ \ref{fig:one} (shown on the leftmost panel) averaged over $100$ trajectories.  Black circles indicate the average distance travelled by each particle after time, from left to right, $t=5 \times 10^1, 5 \times 10^2, 5 \times 10^3, 5 \times 10^4$. At this density, $\tau_\alpha \approx 10^4$.}{fig:propensity}{b}{1}

The models we study are a generalization of constrained lattice gases, first introduced as models of glasses in Ref.\  \cite{Kob-Andersen}. In these models hard-core particles occupy the vertices of a lattice, with single occupancy per site, and with no static interactions between them.  For the dynamics, particles try to hop to neighboring sites, but the local hopping rates depend on the occupancy of surrounding  sites, so as to mimic steric interactions.  These models have the trivial thermodynamics of a non-interacting lattice gas, but their dynamics at high densities displays many of the features of glass formers, such as non-exponential relaxation \cite{Kob-Andersen,Toninelli}, dynamic heterogeneity \cite{Pan}, transport decoupling \cite{Pan}, and aging \cite{Sellitto}  (see Ref.\ \cite{Ritort-Sollich} for a review).

\fig{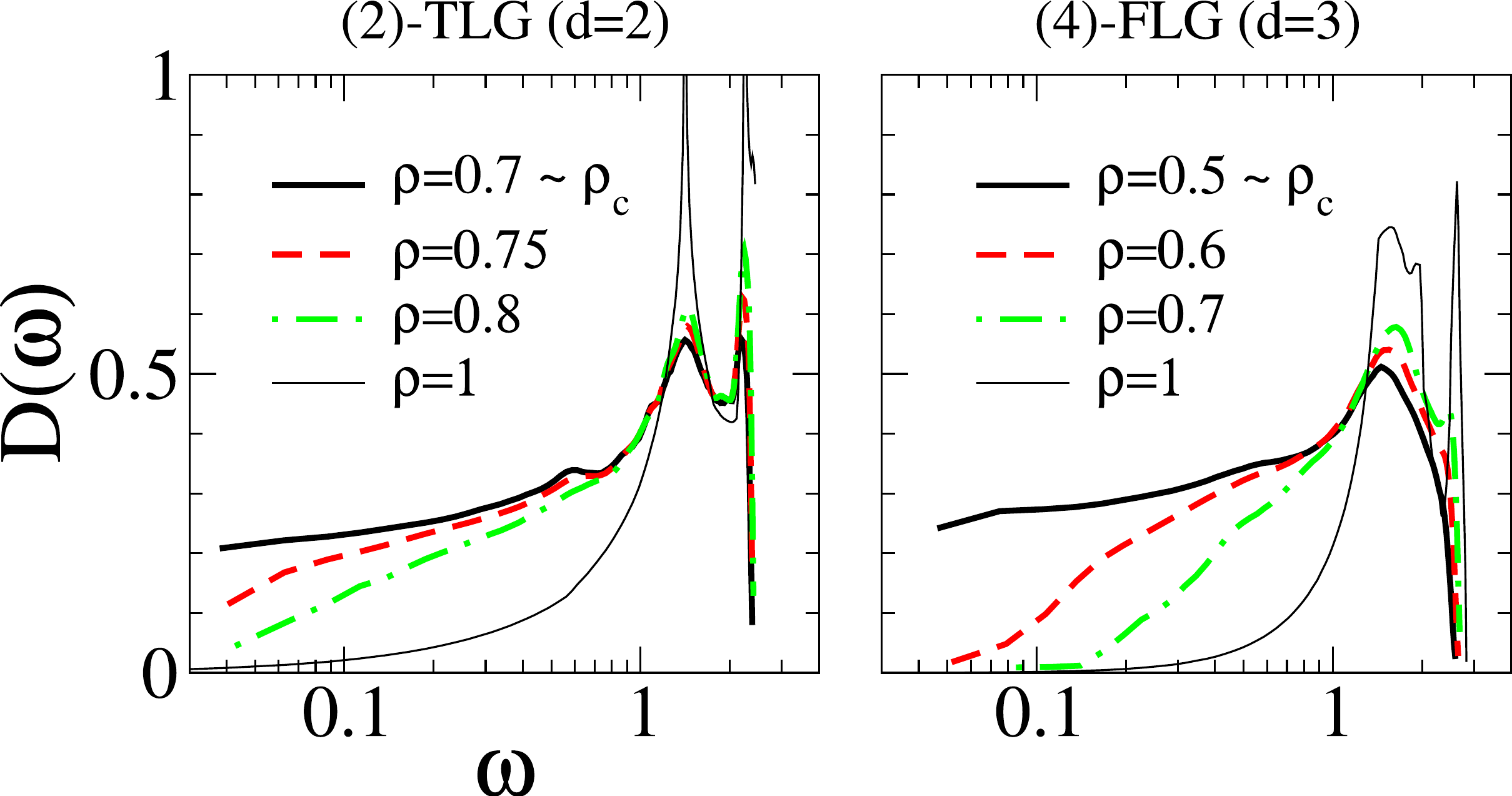}{Density of states for the central force networks of the (2)-TLG model (left panel) and the (4)-FLG model (right panel), at various densities $\rho$ above the isostatic density $\rho_c$.  The curves at $\rho=1$ are exact.  The curves for all other densities were obtained by numerical diagonalization of the dynamic matrix of systems with 4000 normal modes.}{fig:three}{b}{1}

In particular, we focus on the two-vacancy assisted lattice gas model on the triangular lattice \cite{Jackle}, or (2)-TLG, where the kinetic constraint is explicitly due to steric restrictions:
a particle can hop to an empty nearest neighbour site only if the 
common two neighbouring sites are also empty.   We also consider its three dimensional variant on an FCC lattice, where the constraint is that the four common neighbours of the sites undergoing the transition are empty.  We call this model (4)-FLG \cite{Ashton}.  

Figure 2 shows propensity maps for the (2)-TLG, which illustrate the spatial localization of relaxational dynamics in the model.  The leftmost panel is a typical configuration of the model at a density, $\rho=0.75$ in this case, for which relaxation is highly non-exponential and heterogeneous \cite{Pan}.  Panels two to five show the average particle displacement at increasing times in the so-called iso-configurational ensemble \cite{Harrowell2}, i.e., for all trajectories starting from the same initial configuration (the one in the leftmost panel).  For a detailed study of propensity in the (2)-TLG model see Ref.\ \cite{Lester}.

In order to study vibrations together with dynamical heterogeneity we generalize the models above by adding harmonic interactions between neighboring particles.  That is, any two occupied nearest neighbor sites interact through a linear spring of force constant $k$ and of rest length equal to the lattice spacing.   The vibrational Hamiltonian then reads,
\begin{equation} \label{eq:fullHam}
H_{\rm vib} = \sum_{\langle i j \rangle}\frac{1}{2} n_i n_j k \left[ \left( \delta \vec{r}_i - \delta \vec{r_j} \right) \cdot \hat{r}^0_{ij} \right]^2, 
\label{Hvib}
\end{equation}
where $\langle i j \rangle$ means that the sum is over nearest neighbor pairs, 
$n_i=0,1$ indicates whether lattice site $i$ is empty or occupied, $\delta \vec{r}_i$ is the displacement of the particle whose equilibrium position is site $i$, and $\hat{r}^0_{ij}$ is the unit vector between sites $i$ and $j$. Each configuration thus gives rise to a disordered elastic network due to the presence of vacancies in the particle configuration.  This elastic problem is well known, as it corresponds to the ``central force'' problem of rigidity percolation \cite{rigidity}.  In particular, Eq.\ (\ref{Hvib}) for the (2)-TLG and (4)-FLG models correspond to the site diluted triangular and FCC central force networks studied in Ref.\ \cite{Thorpe}.

Figure 3 shows the average density of states (DoS), $D(\omega)$, of vibrational modes for elastic networks corresponding to the (2)-TLG model and (4)-FLG model at various densities.  For densities lower than that of the full lattice, $\rho < 1$, the DoS presents an excess of low frequency modes. Debye scaling, $D(\omega) \propto \omega^{d-1}$, is recovered at low enough frequencies. 
The presence of excess modes becomes more pronounced as the density decreases.  At a critical density $\rho_c$ the excess tail extends all the way to $\omega=0$.  This is the isostatic point at which the system is marginally rigid \cite{rigidity}.  For densities below $\rho_c$ there is an extensive number of zero frequency modes.  We find that for the (2)-TLG $\rho_c \approx 0.7$ and for the (4)-FLG $\rho_c \approx 0.5$, which agrees with the results of Ref.\ \cite{Thorpe} for the triangular and FCC lattices, respectively.

For all densities below that of the full lattice, $\rho < 1$, the vibrational modes are spatially distributed in a non-trivial manner.  Figure 1 illustrates this.  It shows the spatial weight, $\nu_i(\omega)$, of modes of low frequency (Fig.\ 1, top-right) and high frequency (Fig.\ 1, bottom-right) for the (2)-TLG configuration of Fig.\ 2, with $\nu_i(\omega) \equiv \vec{e}_{i\omega} \cdot \vec{e}_{i\omega}$ where $\vec{e}_{i\omega}$ is the two dimensional $i$-th component of the eigenvector of frequency $\omega$.  With this definition $\sum_i \nu_i(\omega) = 1$ for all $\omega$.  Fig.\ 1 (top-right) also shows the propensity map for the same configuration (cf. Fig.\ 2) overlayed on the low frequency eigenvectors.  The picture suggests a close correlation between areas of high propensity for relaxational motion and the location of soft modes.  Fig.\ 1 (bottom-right) shows a similar spatial correlation between regions of low propensity (shown as the inverse of Fig.\ 1, top-left) and the location high frequency modes.

\fig{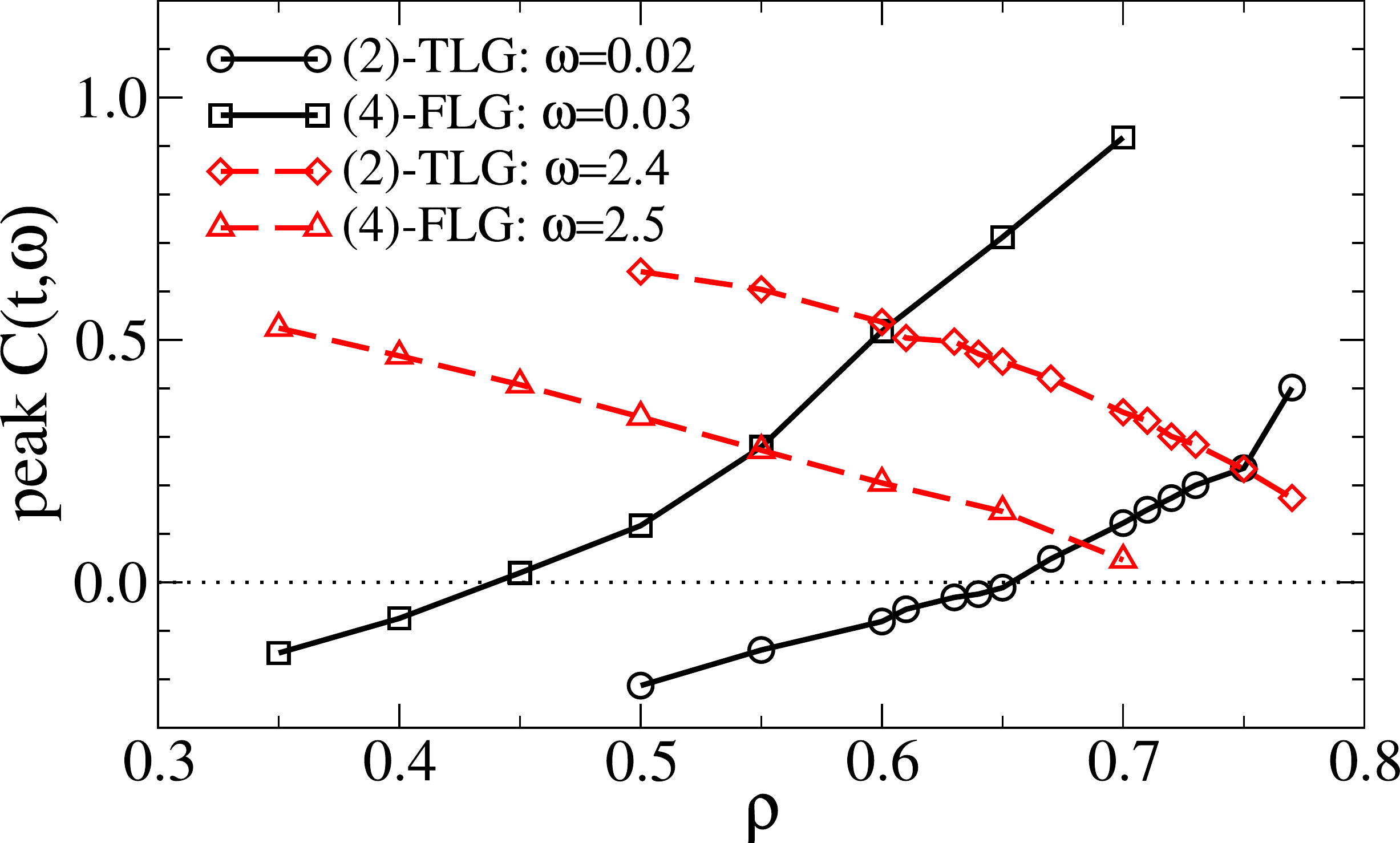}{Peak in the absolute value of the correlation $C(t,\omega)$ between propensity for relaxational motion and vibrational modes, in the (2)-TLG and (4)-FLG, as a function of density; cf. Fig.\ 1 (bottom-left).  The correlation between fast regions/high-frequency modes decreases with increasing density.  The correlation between slow regions/low-frequency modes increases with increasing density.}{fig:four}{b}{.9}

In order to quantify this spatial correlation we define the following cross correlation between local propensity for motion and vibrations:
\begin{equation}
C(t,\omega) = \langle p_{i}^{\mathrm{IC}}(t) \nu_i(\omega) \rangle - \langle p_{i}^{\mathrm{IC}}(t) \rangle .
\label{C}
\end{equation}
Here $p_i(t)$ denotes the persistence function \cite{Pan} of particle $i$, i.e. $p_i(t)=1$ if
particle $i$ has not moved up to time $t$, and $p_i(t)=0$ otherwise.  $p_{i}^{\mathrm{IC}}(t)$ is the iso-configurational average of $p_i(t)$, i.e. the average of the persistence field over all trajectories that start from a given configuration \cite{Lester}.  The average $\langle \cdot \rangle$ in Eq.\ (\ref{C}) is over all equilibrium configurations at a given density.   Fig.\ 1 (bottom-left) shows $C(t,\omega)$ as a function of time $t$ for various vibrational frequencies for the (2)-TLG at density $\rho=0.75$.  For high frequencies the correlation is positive, indicating that particles which are more persistent than average also participate in high frequency vibrational modes. For low frequencies the correlation is negative, indicating that fast relaxing particles (low $p_{i}^{\mathrm{IC}}$) are also those which participate in soft vibrations.  In both cases the correlation is non-monotonic in time, peaking at times around $\tau_\alpha$.

Figure 4 shows how the spatial correlation between vibrations and dynamic heterogeneity depends on density.  It plots the peak value of $C(t,\omega)$ for vibrational modes of the lowest and highest non-trivial frequencies accessible in the simulations.  The fast-relaxation/soft-mode correlation increases with density for all densities larger than the isostatic one, $\rho>\rho_c$.  In contrast, the slow-relaxation/high-frequency correlation decreases with increasing density.  These trends are similar in dimension two, (2)-TLG, and dimension three, (4)-FLG.  Below the isostatic point the correlation between slow regions and low frequency modes changes sign as the vibrational spectrum becomes plagued by zero modes.

The heterogeneity of vibrational modes in the (2)-TLG and (4)-FLG increases with decreasing particle density.  This vibrational heterogeneity can be seen by computing the localization length of each mode.  We do so by using the method of Ref.\ \cite{Nelson}, where a mode localization length is extracted by tracking the change in its eigenvalue due to asymmetric perturbations of the dynamical matrix \cite{Hatano}.  Figure 5 (top) shows the change in the localization length computed in this way for different densities in the (2)-TLG.  In the full lattice limit, $\rho=1$, there are no vibrational anomalies and all the modes are extended.  As soon as a small density of vacancies is present some of the modes become localized, as illustrated in Fig.\ 5 for $\rho=0.95$.  More modes become localized as the density of particles decreases.  Near to the isostatic point most modes have localization lengths that are smaller than our system size, see middle panel of Fig.\ 5 (top).  Beyond the isostatic density the system becomes elastically unstable. 

Elastic heterogeneity also becomes evident in the non-affine response of our system to an external strain \cite{Lubensky,Barrat}.  Following Ref.\ \cite{Barrat}, we deform the system through a uniform strain in the $x$-direction, $\epsilon \ll 1$, by rescaling all particle positions in an affine manner, $\vec{r}_{i0} \to \vec{r}'_{i0}$. After the stretch we minimize the energy using conjugate gradient
descent. Because of the periodic boundary conditions, once the
affine displacement has been made the system will not be able to
relax back to its original configuration. The final positions of
the particles can be expressed as the sum of an affine part,
$\vec{r}_{i0}'$ and a nonaffine part, $\vec{u}_i$.

For high densities the conjugate gradient algorithm is very
efficient at finding the minimum energy configuration.  As density
is lowered towards the rigidity threshold the time to converge starts to increase rapidly.  This is due to the
formation of weakly connected clusters that can make large
displacements at a small energy cost.  These clusters are the
beginnings of the floppy modes that are able to operate almost
independently from the rest of the system.  At the isostatic point
gradient descent cannot converge.  To prevent these divergences we have added a small confining harmonic potential of force constant $\Omega$ at each site to Eq.\ (\ref{Hvib}).  This extra term is diagonal, and so the normal modes remain unchanged but with a shift in frequency.

Figure 5 (bottom) shows the average magnitude of the non-affine response, $\langle (\vec{u}_i/\epsilon)^2 \rangle^{1/2}$, as a function of the density in the (2)-TLG model.  Close to the isostatic point the average non-affine deformation increases very rapidly, and would appear to diverge when $\Omega \to 0$.  The inset to Fig.\ 5 (bottom) has the projection of the non-affine deformation on the normal modes, $\langle u | \omega \rangle \equiv \sum_i \vec{u}_i \cdot \vec{e}_{i\omega}$, as a function of frequency of the modes, for the (2)-TLG at density $\rho=0.75$, showing that non-affinity is carried preferably by softer modes.  

In summary, we have studied the correlation between dynamic heterogeneity and anomalous vibrations in a two and a three dimensional constrained lattice gas model of glasses.  The structural relaxation of these models at high density is similar to that of glass formers, displaying non-exponential relaxation and dynamic heterogeneity \cite{Pan,Lester,Ashton}.  Their vibrational properties are those of well-studied random networks \cite{Thorpe}, and mimic characteristic aspects of the anomalous vibrations of glasses: excess low frequency modes, non-affinity and elastic heterogeneity, all related to the presence of an isostatic point. 
We have found that the location of anomalous vibrational modes correlates to dynamic heterogeneity of structural relaxation, as is observed in atomistic systems \cite{Wyart1,Wyart2,Harrowell}.  In our case, however, structural relaxation, and therefore dynamic heterogeneity, originate in localized vacancies \cite{Toninelli,Lester}, and not in the extended structures that the soft-modes span.  In fact, vacancies act as quenched localized defects for the vibrations, cf. Eq.\ (\ref{Hvib}), giving rise to the anomalous elastic behaviour observed.  We have thus shown through these simple examples that a correlation between soft modes and propensity does not imply a causal relation for relaxation mechanisms.  A similar situation may hold in atomistic models as well.

\

This work was supported by EPSRC grant GR/S54074/01.

\fig{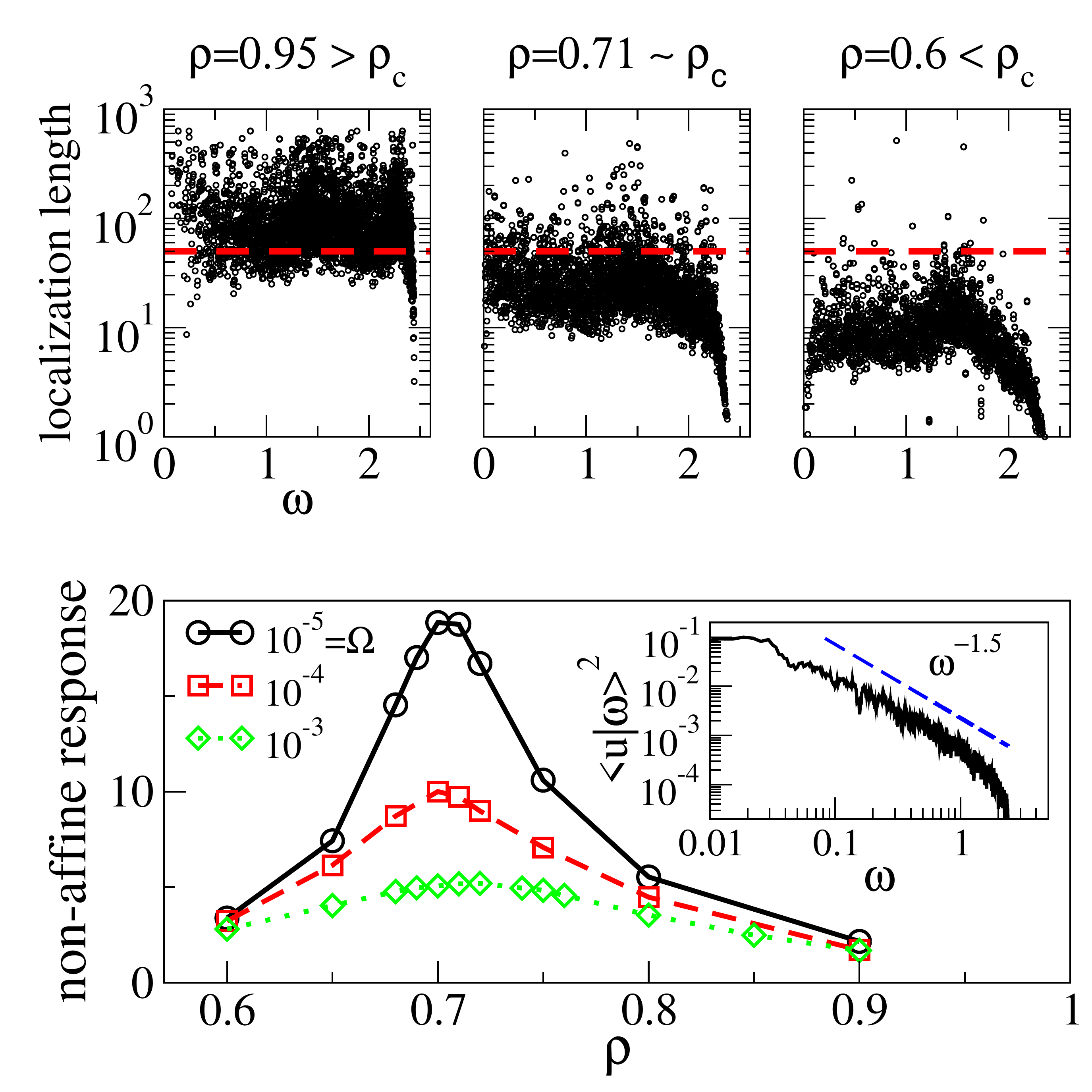}{(Top) Localization length of the vibrational modes of the (2)-TLG,  calculated using the technique or Ref.\ \cite{Nelson}. The dashed line indicates the box length, $L=50$.  As the density decreases towards the isostatic point more modes become localized; for densities below the isostatic one the system breaks up into disconnected elastic components.  (Bottom) Average non-affine response, $\langle (\vec{u}_i/\epsilon)^2 \rangle^{1/2}$, as a function of density in the (2)-TLG.  The different curves are for different strengths $\Omega$ of the confining potential.  In the limit of $\Omega \to 0$, the non-affine response appears to diverge at $\rho_c$.  Inset: Projection of the non-affine displacement into the vibrational eigenmodes, at $\rho=0.75$ for $\Omega=10^{-5}$.}{fig:asymmlocal}{ht}{1}

\end{document}